\documentstyle[prb,aps,eqsecnum,twocolumn]{revtex}

\newcommand{\la}{Landau level}
\newcommand{\beq}{\begin{equation}}
\newcommand{\eeq}{\end{equation}}
\newcommand{\bea}{\begin{eqnarray}}
\newcommand{\eea}{\end{eqnarray}}
\newcommand{\qy}{q_{y}}
\newcommand{\qx}{q_{x}}
\newcommand{\Qx}{Q_{x}}

\newcommand{\qiy}{q_{iy}}

\newcommand{\qoy}{q_{1y}}

\newcommand{\qty}{q_{2y}}
\newcommand{\Qi}{Q_{i}}

\newcommand{\pri}{^{\prime}}

\newcommand{\rkqy}{\rho_{k,k+\qy}}
\newcommand{\rkkp}{\rho_{k,k^{\prime}}}
\newcommand{\rrrp}{\rho ({\bf r},{\bf r^{\prime}})}

\newcommand{\mml}{M(m,l)}
\newcommand{\sik}{\Sigma_k}
\newcommand{\emm}{(\epsilon - \mu)}
\newcommand{\nq}{{\tilde{n}}({\bf{q}})}
\newcommand{\vft}{{\tilde{V}}}
\newcommand{\uh}{{\tilde{u}}_{{\rm H}}(p)}
\newcommand{\uf}{{\tilde{u}}_{{\rm F}}(p)}
\newcommand{\uhf}{{\tilde{u}}_{{\rm HF}}(p)}
\newcommand{\vag}{{\tilde{V}}_{{\rm AG}}}
\newcommand{\eag}{\epsilon_{{\rm AG}}}

\newcommand{\ix}{I_x}
\newcommand{\iot}{I_{12}}

\newcommand{\Dp}{d^{2}{\bf{p}}}
\newcommand{\Dq}{d^{2}{\bf{q}}}
\newcommand{\Dr}{d^{2}{\bf{r}}}

\newcommand{\wq}{\omega_{n}({q^2\over{2}})}
\newcommand{\wqt}{\omega_{n}({q^2/{2}})}

\newcommand{\aq}{\alpha_{n}({q^2 \over{2}})}
\newcommand{\aqt}{\alpha_{n}({q^2/{2}})}
\newcommand{\ape}{\alpha_{n}({p^2 \over{2}})}
\newcommand{\apet}{\alpha_{n}({p^2/{2}})}
\newcommand{\nt}{$n ^{\rm{th}}$ }

\newcommand{\iqr}{^{i{\bf{q \cdot r}}}}
\newcommand{\br}{{\bf{r}}}
\newcommand{\bq}{{\bf{q}}}

\newcommand{\rla}{{\rm{R_{L}}}}
\newcommand{\kfe}{k_{\rm{F}}}
\newcommand{\ngi}{$n \rightarrow \infty$}

\begin{document}
\title{Exact results for interacting electrons in high Landau levels}
\author{R. Moessner and J. T. Chalker}

\address{Theoretical Physics, Oxford University,
1 Keble Road, Oxford OX1 3NP, UK}
\date{\today}
\maketitle
\begin{abstract}
We study a two-dimensional electron system in a magnetic
field with a
fermion hardcore interaction and without disorder. 
Projecting the Hamiltonian onto the $n^{th}$~\la, we show
 that
the Hartree-Fock theory is exact in the limit \ngi, for the high 
temperature, uniform density phase of an infinite system; for
a finite-size system, it is exact at all temperatures.
In addition, we show that a
charge-density wave arises below a transition temperature $T_t$. Using
 Landau
theory, we construct a phase diagram which contains both
unidirectional and 
triangular charge-density  wave phases. We discuss the unidirectional
charge-density wave at zero 
temperature and argue that quantum fluctuations are unimportant in the
large n limit. Finally, we discuss the accuracy of the
Hartree-Fock approximation for potentials with a nonzero range such as
the Coulomb interaction.

\end{abstract}


\input{psfig}

\section{INTRODUCTION}

In a two-dimensional electron system (2DES) in a strong magnetic
field, the fractional quantum
Hall effect [\onlinecite{fqhe}] may be observed when electrons
partially 
fill the lowest ($n = 0$) \la. In this paper, we consider what happens
if it is 
a high \la~that is partially occupied. 
The large-$n$~limit generates a problem that we can, in part, solve
exactly. To this extent, the high \la~limit is analogous to the limit
of high dimensions for spin models
[\onlinecite{brout}] or the
 Hubbard model [\onlinecite{voll}].

Electron states in lower excited \la s ($n = 1,2$) have received some
attention in the past 
[\onlinecite{old}], and recently
Aleiner and Glazman
[\onlinecite{ag}] have derived an effective interaction for electrons
in high \la s, taking into account the wavevector dependent dielectric 
constant due to screening by lower, fully occupied levels.
 Using this
interaction, Fogler, Koulakov and Shklovskii (FKS) [\onlinecite{kfs}]
have 
studied a 2DES in a weak magnetic field at zero temperature, within
the 
framework of the Hartree-Fock approximation. For a filling fraction, 
$\nu$, of the partially filled \la~near $\nu = 1/2$, they find 
 a phase of fully occupied (incompressible)  regions alternating
with regions of 
unoccupied orbitals. These regions are translationally invariant in
one dimension ('stripes'). At lower filling fractions, these stripes
break up and there is a phase of fully occupied islands ('bubbles') on
a triangular 
lattice in an otherwise empty  system. The width of the
stripes and the lattice constant are each of order of the Larmor
radius. Experimental consequences of this structure, in particular for
tunneling into the 2DES, have been investigated extensively by
FKS.  

Both the interest and the difficulty of studying electron correlations
in a partially filled \la~stem from the fact that the single-particle
problem is macroscopically degenerate. Interactions set the only
important energy scale and the problem is intrinsically
non-perturbative. 
In the $n^{th}$~\la,
however, a small parameter does appear, which manifests itself in the
existence of two new lengthscales besides the magnetic length, $l =
\sqrt{\hbar/(eB)}$: $\lambda_{{\rm F}}= {2 \pi /{\kfe}} = {2
\pi l/{\sqrt{n}}}$, the Fermi wavelength, and the Larmor radius,
$\rla \approx l 
\sqrt{2n}$, the size of the smallest wavepacket that can be
constructed in the $n^{th}$ \la. We use this small parameter,
$n^{-1}$, 
to obtain a problem that is
amenable to exact solution.

We consider a 2DES assuming that the partially filled \la~is
spin-polarized
 Using a fermion hardcore
potential of strength $v \ll \hbar \omega_c$ as a model interaction
between  the electrons, we are able to
resum  perturbation theory at high temperatures, for
\ngi. ($\hbar \omega_c = \hbar {eB/{m}}$~is the cyclotron
energy and from hereon, we set $l \equiv \hbar \equiv 1$). 

We find that, for the
high-temperature phase, the Hartree-Fock 
approximation is exact in the large-$n$ limit (section
\ref{sect.diagram});  for a system 
on a cylinder of finite circumference, it remains exact down to zero
temperature (section \ref{subsect.vertex}) 
and the 
eigenstates are single Slater determinants of Landau orbitals.   
The high temperature phase is unstable towards the formation of a
charge-density wave (CDW) of wavelength $\lambda_{{\rm min}} \simeq
2.9 \rla$ (section \ref{subsect.bseqn}). We construct a phase diagram
for the system  
using a Landau expansion (section \ref{sect.cdw}).
 We find that
there exist both a unidirectional CDW, near $\nu = 1/2$, and a
triangular CDW away from $\nu= 1/2$.

Our work hence generalises to finite temperatures the results obtained
by FKS  at zero
temperature. It also represents a generalisation of the study of
charge-density waves in the lowest \la~by Fukuyama, Platzmann and
Anderson [\onlinecite{fpa}]. In particular, we show 
that the transition from the uniform state near $\nu = 1/2$~is to a
unidirectional rather than triangular CDW. 

We also (section \ref{sect.finiterange}) compare the theories with
hardcore 
and finite range potentials.
It emerges that the Hartree-Fock approximation is {\em not} exact for
potentials of finite range $R$~(section \ref{subsect.nonhf}) even in the
large-$n$~limit. 
We show, however, that corrections to the Hartree-Fock approximation
are governed by the parameter $R/l$, which is small in the weak-field
limit investigated by FKS. For potentials where $R/l$~is not
small,  
we demonstrate explicitly that there exist
uniform density states which, for these interactions, have a cohesive
energy of the same order as that of the CDW states. It therefore
remains possible that the groundstate in high \la s may
depend on details of the interaction, and may not necessarily have CDW
order.


\section{The Hamiltonian}

We start by considering the Hamiltonian projected onto one spin
orientation for the \nt \la:
\bea
H & = & {1\over{2}} {\displaystyle \sum_{r,q,p,k}} \langle r q|V|p k
\rangle a_{r}^{\dagger} 
a_{q}^{\dagger} a_{k} a_{p} \nonumber \\
 & = & {1\over{2}} \sum_{k,m} \int dl \int{\Dq \over{(2 \pi)^2}}~
\vft(\bq)~ \aq~e^{-i\qx m}  \times \nonumber\\
&& \delta (\qy -l) a_{k+l}^{\dagger}
a_{k+m}^{\dagger}  a_{k+m+l} a_{k}
\label{eqn.hami} \\
 & \equiv &  {1\over{2}} \sum_{k,m,l} M(m,l)~a_{k+l}^{\dagger}
a_{k+m}^{\dagger} a_{k+m+l} a_{k}. 
\eea
Here, $a_{p}^{\dagger}$ creates an
electron in the Landau orbital $\langle {\bf r}|p \rangle =
\psi^{(n)}_{p}({\bf r}) $ of
the \nt \la~with
pseudomomentum $p$, $\psi^{(n)}_{p}(\br) = L^{-{1\over{2}}}
e^{ipx}~\phi^{(n)} (x-p)$, where
$\phi^{(n)}$ is the \nt harmonic oscillator wavefunction.
Periodic boundary conditions are applied in the y-direction for a
system size $L$. Also, $\aqt = (\wqt)^2 = 
({\rm{L_{n}}}({q^2/{2}}) e^{-{q^2/{4}}})^2$, ${\rm{L_{n}}}$
being the \nt Laguerre polynomial.
The Fourier
transform of the potential is $\vft(\bq)=\int{\Dr}~e^{i{\bf{q \cdot
r}}}~V(\br)$, and all quantities $\tilde{f}$~with a tilde are to be
understood as the Fourier transform of the function $f$. 
To derive (\ref{eqn.hami}), we have used [\onlinecite{rs}] 
\beq
\label{eqn.mxelt}
\langle k \pri |e \iqr |k
\rangle = {2 \pi\over{L}} \delta (k \pri -k-\qy)e^{({i\qx(k+k
\pri))/2}}~\wq .
\eeq
We represent the vertex $\mml$~, which is independent of k,
as in Fig. \ref{f1}.  

In this paper, we consider primarily the simplest possible potential:
a contact interaction for fermions. In the lowest \la~this interaction
is known to reproduce much of the physics of the fractional quantum
Hall effect [\onlinecite{macdrev}] and the Laughlin wavefunction
[\onlinecite{lau}] is
known to be its exact [\onlinecite{tk}]
ground state at $\nu = 1/3$. 
It is highly singular at the origin in real
space but has a well-behaved Fourier transform

\beq 
\tilde{V}_{{\rm HC}}(q) = -v q^2,
\eeq
$v$ being the interaction strength.

In addition, in section \ref{sect.finiterange} we discuss the screened
Coulomb interaction, $V_{{\rm AG}}$, recently derived by Aleiner and
Glazman [\onlinecite{ag}], as well as two simple finite
range potentials, $V_1$~and $V_2$.
The potential $V_{{\rm AG}}$~(\ref{eqn.vag}) is an effective
intra-\la~interaction potential 
which takes into account the presence of
lower, fully-occupied \la s through the dielectric constant $\eag$. It 
has the following important features: there is little screening
for $q \alt n^{-1}$~and $q \agt 1$, but for $q \sim
n^{-{1\over{2}}}$, $\eag \sim \rla / a_{{\rm B}}$. In weak magnetic
fields, this is very large, of order $n^{1\over{2}}$, whereas in
strong  fields, $\eag \approx 1$~for all $q$. The explicit form of
$V_{{\rm AG}}$~is  

\beq
\vag (q) = {2 \pi e^2\over{4 \pi \epsilon_0 \kappa \eag
(q) q}},
\label{eqn.vag}
\eeq
where 
\beq
 \eag (q)= 1 + {2\over{q a_B}}
\left[1-J_0^2(q \rla) 
\right],
\nonumber
\eeq
$J_0$~is the Bessel function, $a_B = {\hbar ^2 4 \pi \epsilon_0 \kappa
/(m 
e^2)}$~is the Bohr radius,  and $\kappa$~is the relative permittivity. 

The two finite-range potentials we consider are of range $R$. 
One  potential is 
box-like, non-zero and constant up to a
distance $R$ and zero for larger distances; the other one decays
exponentially with a decay length
$R$. We take $\lambda_{{\rm F}} \ll R \ll \rla$. The Fourier
transforms  of these potentials are
\beq
\vft _1(p) = v_1
R^2 J_1(qR)/qR;\hspace{0.5cm} \vft _2(p) = v_2 R^2 (1+
q^2\:R^2)^{-{3\over{2}}}
\label{eqn.finiterange}
\eeq
where the $v_i$~are the strengths.
  Potentials of a finite range arise, for example, if there
is a metallic gate close to the two-dimensional electron system which
effectively screens the interaction at distances larger than the
2DES-gate separation.


\section{The limit \ngi}


\subsection{The vertex $\mml$}
\label{subsect.vertex}

We first investigate qualitatively the behaviour of the individual
matrix elements. To do so, we replace $\aqt$~ by its WKB envelope
[\onlinecite{bo}], normalized to be consistent with
$\int_{0}^{\infty} ({\rm L_{N}}(x))^2 e^{-x} dx = 1$.  We set $\aqt =
0$~for $q^2 > 8n$, and for $q^2 < 8n$
\beq
\aq = {2\over{\pi}} {1\over{\sqrt{q^2(8n-q^2)}}}.
\label{eqn.wkb}
\eeq
This is an asymptotic expression away from $q=0$ and $q=\sqrt{8n}$. In
the vicinity of these points, we have used the following expressions
[\onlinecite{erd}]:
\bea
\aq & = & J_0^2(q \sqrt{2n}) \hspace{2.2cm} q^2  \ll  n
\label{eqn.wkbturnz} \\ 
\aq & = & (\Gamma({2\over{3}})18^{1\over{3}})^{-2}
n^{-{2\over{3}}} \hspace{0.7cm}  q^2 - 8n \ll n^{1\over{3}}
\label{eqn.wkbturnn}
\eea
In considering the envelope only, we neglect the 
rapid oscillations of $\aqt$, since the potential does not have
structure on this scale.

To estimate $\mml$~ for the hardcore potential we rescale $\qx =
\sqrt{8n}\:\Qx$~and find 
\bea
\mml & = & {2 \pi \over{L}} 2 \int_{0}^{\sqrt{1-{l^2\over{8n}}}} {d      
\Qx\over{2\pi}} \sqrt{8n} 
(-v (\Qx^2+{l^2\over{8n}})) \times \nonumber \\
&& {\rm cos}(\Qx
m\sqrt{8n}){1\over{\sqrt{\Qx^2(1-(\Qx^2+{l^2\over{8n}}))}}}. 
\eea
For large $n$, the oscillating cosine suppresses all
$\mml$~ for which $m \gg n^{-{1\over{2}}}$.

The consequences of these simplifications are particularly clear if we
consider a system on a cylinder of finite circumference $L$. Then,
$m=2 \pi
\alpha /L$, with $\alpha$~integer, so that, in the limit \ngi~with
$L$~fixed, only $M(m=0,l)$ is non-zero [\onlinecite{turning}]. 
 The  Hamiltonian in this limit thus consists only of exchange terms
$(m=0)$. It is therefore 
diagonal in a basis of Slater determinants of Landau orbitals.
 The charge density of any single such Slater determinant
is independent of the y-coordinate. 
The problem of finding the ground state is therefore
reduced to finding the unidirectional charge-density wave which is
lowest in energy, and the Hartree-Fock approximation for this system
is exact.

Note that the spatial dependence of the exchange
interaction is as follows: 
$M(m=0,l)$~varies little
as a function of $l$ for $|l| < 2 \rla$. For $|l| > 2 \rla$, the
cyclotron orbits do not overlap.
The effective range of the contact interaction is
therefore determined by the spatial extent of the wavefunctions and is
thus of order of the Larmor radius.

It is interesting to compare our conclusions with those of 
Rezayi and Haldane [\onlinecite{rh}], who considered a similar problem
in the lowest \la. They found by numerical diagonalisation that a
single Slater determinant of Landau orbitals is the ground state
if the magnetic length is much larger than the circumference of the
cylinder. In the system we consider, the important lengthscale is
$\rla$~and the limiting behaviour is reached for $L \ll \rla$.


\subsection{Diagrammatic analysis}
\label{sect.diagram}
\label{subsect.diaasym}

We now turn to the infinite system and show by means of a
diagrammatic analysis that, in the absence of symmetry breaking, the
Hartree-Fock approximation is exact for the hardcore
potential, in the high temperature phase, in the limit \ngi. To do
this, we group 
all diagrams for the self-energy into three classes (Fig.
\ref{f15}): first, those 
diagrams which contain neither closed fermion loops nor crossed
interaction lines (class A); second
those which do
not contain closed fermion loops but do contain crossed interaction
lines  (class B); and third
those which contain closed fermion loops (class C).

For the analysis of the $n$-dependence of these diagrams, we note
that the sum over Matsubara frequencies and the integral over
intermediate states decouple: the bare propagators, 
$G^{(0)}_{\omega_{m},k}$, which are diagonal
in the pseudomomentum $k$, are the same for all states in a given \la
, since 
$G^{(0)}_{\omega_{n},k} = (i 
\omega_m - (\epsilon_{n}(k)-\mu))^{-1}=(i\omega_m - (\epsilon -
\mu))^{-1}$~. There is no $n$-dependence in the propagators since the
dependence on $n$~in  $\epsilon_{n}$~ is absorbed
into an $n$-dependent chemical potential $\mu$. Therefore, only the
behaviour of the integrals over intermediate states has to be
considered when taking the large-$n$~limit.

For a diagram in the first class with one interaction line, the
integral 
over intermediate states is 
\beq
\sum_{\qy} M(0,\qy) = -{8 v n\over{\pi^2}}\int_{0}^1 {Q^2
dQ\over{\sqrt{1-Q^2}}} = -{2 v n\over{\pi}}
\label{eqn.norm}
\eeq

This class of diagrams in fact makes the leading contribution to
the self energy for \ngi~
and hence we scale $v$ with $n$
so that  $vn$ is constant.
Next we sum all the  diagrams
in this class; in appendix \ref{app.dia} we establish an upper bound
of $(vn)^x n^{-{2\over{3}}}$ to the contribution from other diagrams,  
which vanishes  for \ngi. The sole exception are the diagrams which
contain only 
Hartree loops (Fig. \ref{f4}) and interaction lines which do not cross;
these vanish 
for our  potential since $\tilde{V}_{{\rm
HC}}(0)=0$. If 
this were not the case, we would have to compare their contribution
with that of the Fock diagrams when considering what to retain for
large $n$. 


\subsection{The self energy $\sik$}
\label{subsect.selfenergy}

We sum the diagrams which do not vanish as $n \rightarrow
\infty$ using the Dyson equation (Fig. \ref{f5}), noting that the
absence of a 
Hartree contribution reduces it to the one in Fig. \ref{f6} so that

\bea
\sik & = & \sum_{l,n_l} {-M(0,l) \over{\beta}} {e^{i E_{n_l} 0^+}
\over{i E_{n_l} - \emm}} \times \nonumber \\
&&\sum_p \left[ \sum_{m, n_m} {-M(0,m)
\over{\beta}} \left( {-1 \over{i E_{n_m} - \emm}} \right) ^2 \right]^p
\nonumber \\
 & = &  {-\nu \sum_{l= - \infty}^{\infty} M(0,l) \over{1- {\beta \nu
(1-\nu ) \sum_{l= - \infty}^{\infty} M(0,l)}}}
\eea
For the hardcore potential,  $\sum_{l=- \infty}^{\infty} M(0,l)$~is
negative (Eq. \ref{eqn.norm}), and the
self-energy has no poles. 

We
see that $\sik$~ is independent of $k$, because of the translational
invariance of the system; to keep the filling fraction $\nu$~fixed,
$\sik$~will be compensated by an equal shift in the chemical
potential. Hence, to leading order in $n$, the one-particle propagator
remains unchanged.


\subsection{The Bethe-Salpeter equation}
\label{subsect.bseqn}

The two-particle vertex $\Gamma$, on the other hand, is affected by
the 
interaction. An analysis like the one above applies also to the
perturbation expansion of $\Gamma$, and in the end we only have to sum
ladder diagrams illustrated in Fig. \ref{f7} [\onlinecite{ladder}].
The Bethe-Salpeter equation (Fig. \ref{f8}), in the limit \ngi, is

\bea
\Gamma (q, \omega ) & = & M(0,q) - \sum_{l,\lambda}
{M(0,l)\over{\beta}} \times \nonumber\\ 
&&{-1\over{i(\omega + \lambda) - \emm}} {-1\over{i(\lambda) - \emm}}
\Gamma(q-l, \omega). \nonumber
\eea
Note that, because the constituent vertex $\mml$~and the propagators
are independent of the  pseudomomentum $k$ and the
Matsubara frequencies $ \Omega$~and $\delta$, so is $\Gamma$.

For $\omega \neq 0$~ the sum over $\lambda$ yields 0 and $\Gamma
(q,\omega ) =
M(0,q)$. For $\omega = 0$

\bea
\Gamma (q,0) & = & M(0,q) + \nonumber \\
&&\beta \nu (1-\nu ) \sum_l M(0,l)  \Gamma
(q-l,0) \nonumber \\
{\rm and} \hspace {1.0cm} {\tilde{\Gamma}}(s) & = &{{\tilde{M}}(0,s)
\over{1- 
{\tilde{M}}(0,s) \beta \nu (1-\nu )}} 
\label{eqn.vertex}
\eea
where ${\tilde{\Gamma}}$~and ${{\tilde{M}}}$~are Fourier transforms of
$\Gamma (q,0)$~and $M(0,q)$~with respect to $q$. $2 \pi
\tilde{M}$~is in fact the Fock potential (\ref{eqn.uf}).

The two
particle vertex is divergent at a temperature $T = {\tilde{M}}(0,s)
\nu 
(1-\nu )$.  This indicates an instability
towards the formation of a charge-density wave at a wavevector
$p_{{\rm min}}$~where
${\tilde{M}}(0,p_{{\rm min}})$ is largest. 
 We discuss the phase diagram of the
charge-density waves in section \ref{sect.cdw} . Had we chosen the
potential to be attractive, the 
system would
phase-separate into a fully occupied and an empty
region below the critical temperature.


\section{Charge-density wave instability and the phase diagram}
\label{sect.cdw}

In the preceeding section, 
we have shown that the uniform system is unstable to formation of
a CDW with a wavevector $p_{{\rm
min}}$~at a temperature $T_2 = \nu (1-\nu) {U}(p_{{\rm min}})$,
with ${U}(p) 
={\uf/(2 \pi)}$. In section
\ref{subsect.selfconsistent}, we find for the hardcore interaction
that $p_{{\rm min}} \simeq
2.16/\rla$. 
For a general potential, replacing $ \tilde{u}_{{\rm F}}$~by $
- \tilde{u}_{{\rm HF}} = - \tilde{u}_{{\rm H}}+ \tilde{u}_{{\rm
F}}$~(\ref{eqn.uh}) in these and subsequent expressions yields a 
general analysis of  
the CDW phase 
diagram for a Hamiltonian in the 
Hartree-Fock approximation. 

The problem of charge-density waves in the lowest \la~has been studied 
extensively [\onlinecite{yf}], and Fukuyama,
Platzmann and  Anderson
(FPA) 
[\onlinecite{fpa}] have obtained the same expression for the
transition temperature using second order perturbation theory with a
Hartree-Fock Hamiltonian. These authors also show that, within Hartree
Fock theory, a second order transition at $T_2$~is preempted by a
first-order transition at a higher temperature $T_1$~to a triangular
CDW. 

In the following, we extend the work by FPA to the
$n^{th}$~\la [\onlinecite{cont}]. In the spirit of Landau theory, FPA
expand the free 
energy in powers of the order parameter $\delta_1$~for the triangular
phase using a
diagrammatic calculation described in 
Refs. [\onlinecite{mr}] and obtain

\beq
\delta F_1  =  N U(Q)(a|\delta_1 |^2 + b |\delta_1 |^3 + c |\delta_1 
|^4 ) 
\nonumber 
\eeq
where the coefficients are given by
\bea
a & = & 3(1-\nu (1-\nu)u) \nonumber \\
b & = & -2 u^2 \nu (1-\nu) |1-2\nu| {\rm cos}({\sqrt{3}\over{4}}Q^2)
\nonumber  \\
c & = & u^3 \nu (1-\nu) \left[  {9 \over{2}}
\left[{5\over{12}} - (\nu - 
{1\over{2}})^2 \right] + \right. \nonumber \\ && \left. \left[ 1 -
{\rm cos}({\sqrt{3}\over{2}}Q^2)\right]\left[6(\nu - 
{1\over{2}})^2 - {1\over{2}}\right] \right].  
\label{eqn.c}
\eea
Here, $Q$~is the wavevector of the triangular phase determined below,
$u = U(Q)/T$~and  
$N$~is the number of states in the 
system. By the same method, we find the free energy for the
unidirectional CDW:
\bea
\delta F_2 & = & N U(Q)(A|\delta_2 |^2 + C |\delta_2 |^4 ) \nonumber  
\eea
where the coefficients are
\beq
A  =  a/3 \; {\rm and} \; 
C  =  u^3 \nu (1-\nu) \left[{1\over{8}} + {1\over{2}}(\nu -
{1\over{2}})^2 \right]. \nonumber
\eeq

The temperature at which the first order transition occurs from the
uniform phase to a triangular CDW is 
\beq
T_1 = T_2 {U(Q)\over{U(p_{{\rm min}})}} \left( 1 + {e\:
{\rm cos}^2({\sqrt{3}\over{4}}Q^2)   \over{f + g\:
{\rm cos}^2({\sqrt{3}\over{4}}Q^2)}} \right), 
\label{eqn.t1} 
\eeq
where 
$ e = {8\over{27}}(\nu -{1\over{2}})^2, f= {7\over{36}}+
{5\over{3}}(\nu -{1\over{2}})^2, g = {2\over{9}} - {8\over{3}}(\nu
-{1\over{2}})^2 \nonumber$.
$T_1$~ is found by maximizing (\ref{eqn.t1}) with respect to
$Q$. For high \la s, this calculation simplifies as follows. Since $Q 
\sim p_{{\rm min}} \sim
n^{-{1\over{2}}}, {\cos}^2({\sqrt{3}\over{4}}Q^2) \simeq 1$. Hence,
for large $n$, $T_1$~is obtained by setting $Q=p_{{\rm min}}$.
Therefore,  $T_1 =
T_2 \left( 1 + {e/(f+g)} \right)$~is always larger than $T_2$,
except for $\nu = {1/{2}}$, where there is a triple
point:$\: T_1=T_2\equiv T_t$.
In the remainder of the calculation, we can set
${\rm cos}^2({\sqrt{3}\over{4}}Q^2) = 1$~and $U(Q) = U(p_{{\rm
min}})$,  thereby
removing the second term in brackets in (\ref{eqn.c}).  

We now turn to the question of which phase is present for $T  
< 
T_t$. 
 Since our expression for the free energy 
is valid only for small $\delta$,  we
consider the neighbourhood of the triple point. Minimizing $\delta
F_{1,2}$~with  respect to
$\delta_{1,2}$, we find that the unidirectional CDW is more
favourable at $\nu = {1/{2}}$~for $T$~just below $T_t$. As the
filling fraction is lowered at constant $T$,
there is a first-order phase transition to the triangular CDW.
When $\nu$ is lowered even further, the 
uniform density state becomes more favourable.
We have also examined competition between the striped phase and a
square CDW, using the expression derived by Gerhardts and Kuramoto
[\onlinecite{yf}];
we find that at $\nu = 1/2$~for $T$~just below $T_t$, the square CDW
has a higher free energy than  the unidirectional one for \ngi.
 The resulting phase diagram is  
displayed in Fig. \ref{f13} for $\nu \leq 1/2$. The results for $\nu >
1/2$~follow from particle-hole symmetry. The emergence of the
unidirectional 
CDW can be explained from the expansion of the free energy as
follows. As $\nu$~approaches $1/2$, $b$~tends to zero. The gain of
$\delta F_1$~relative to $\delta F_2$~due to its larger second-order
coefficient, $a = 3A$, which is almost zero near $T_t$, is 
offset by the cost incurred due to its larger fourth-order
coefficient $c > C$, and the unidirectional CDW is energetically
more favourable. 

As the temperature is lowered far below $T_1$, the order parameter
grows and our expansion ceases to be valid. An additional difficulty
is that higher harmonics will start to play a role: 
the density profile must be constructed  
 by occupying states in the $n^{th}$~\la, and must in particular
not require a filling fraction larger than one at any point. In
section 
\ref{subsect.selfconsistent} we show that the cost for higher
harmonics is small. We therefore expect that these
considerations alter only the shape of the CDW without changing 
its 
structure or periodicity. In this way, we expect competing
unidirectional  and
triangular patterns to persist at wavevector $p_{{\rm min}}$~down to 
zero temperature. 

If we extrapolate the phase-separation line $T_{12}$~beyond the
range of 
validity of our expansion to $T=0$, we obtain a phase
diagram (Fig. \ref{f14})  with a 
first-order 
transition at $\nu = 0.35$~for $T=0$.  This is close to the
value  $\nu = 0.39$ obtained numerically by FKS, separating the
unidirectional pattern from the triangular one.


\section{Zero temperature analysis}

\subsection{The effective Hartree potential}
\label{sect.effhf}

We start this section with a  general remark concerning  the
Hartree-Fock 
approximation for a Hamiltonian projected onto a single \la.
MacDonald and Girvin [\onlinecite{mg}] have shown, using analyticity
in  
the  lowest \la, that the Fock
potential can be expressed solely in terms of  
electron density as a function of position.
This result is in fact
not a consequence of the special analyticity properties of the
wavefunctions in the lowest \la : if the Hamiltonian is projected onto
any 
single \la, the Hartree Fock energy is simply $1/2 \int \Dq
\tilde{V}_{{\rm eff}}({\bf q}) |\nq |^2$, where $\nq$~is the Fourier
transform of the electron density and the effective potential is

\bea
{\tilde{V}}_{\rm{eff}}({\bf q}) & = & {\tilde{V}}({\bf q}) -
{\tilde{V}}^{\rm{F}}_{\rm{eff}}({\bf q}) 
\label{eqn.hartreeeff} \\
{\tilde{V}}^{{\rm{F}}}_{{\rm{eff}}}({\bf q}) & = &
{{\tilde{u}}_{{\rm{HF}}}({\bf{q}}) \over{\aqt}} \nonumber\\
& = & {1 \over{\aqt}}
\int 
{\Dp \over{2 \pi}} {\tilde{V}}({\bf{p}}) e^{-i{\bf{p \cdot
q}}} \ape
\label{eqn.fockeff}
\eea
The crucial point is that after projection, the density matrix in the
Landau orbital 
basis, $\rkkp$ can be expressed in  terms of  $\nq$. Details of the
derivation of 
(\ref{eqn.fockeff}) are given in
appendix \ref{app.fock}.


\subsection{Self-consistent equations}
\label{subsect.selfconsistent}

The diagrammatic analysis of section \ref{sect.diagram}~applies to the
high temperature phase where the electron liquid has a uniform
density. For that case, the Hartree-Fock
approximation is exact as $n \rightarrow \infty$. We now allow for
symmetry breaking and solve the self-consistent set of equations
depicted in Fig. \ref{f5}. For a system on a cylinder of finite
circumference, we have shown 
that the Hartree-Fock eigenstates have a probability density that is
translationally  invariant in the
y-direction and hence, that it is sufficient to allow the self-energy
to depend on one co-ordinate only. 
(The infinite system is discussed further in section
\ref{subsect.stability}). The Dyson equation reads

\bea
\sik & = & {1 \over{2 \pi}} \sum_l \int dE \left[-M(l-k,0) {i \over{E
- \sik -\emm}} \right. + \nonumber \\
& & \left. M(0,l-k) {i \over{E- \sik -\emm}} \right] \nonumber  
\\ 
 & = & \sum_l [M(l-k,0)-M(0,l-k)] \langle n_l \rangle \nonumber
\eea
where $\langle n_k \rangle = -\Theta( \sik + \emm)$~is the occupancy
of the 
Landau 
orbital $|k \rangle$~and $\Theta$~is the Heaviside step function. This
gives 

\bea
\tilde{\Sigma}(p) & = & {1\over{2 \pi}} \uhf \tilde{n}(p), 
\label{eqn.selfenergy} \\
{\rm where} \;
\uh & = &\tilde{V}(p) \ape =  -(vn) {p^2 \ape \over{n}},
\label{eqn.uh} \\
- \uf & = & - \int q dq \vft (q) \aq J_0(pq) 
\label{eqn.uf} \\
 & \simeq & 4(vn)\: _{1}F_{2}({3\over{2}};1,2;-2np^2), 
\label{eqn.ufas}
\eea
$_{1}F_{2}$~being a hypergeometric function.

Note that the Fock potential (\ref{eqn.uf}) follows directly from 
(\ref{eqn.fockeff}) as it must. The factor of
$\apet$~arises from 
the matrix elements of the potential between states in the $n^{th}$
\la. Eqs. (\ref{eqn.uh}) and (\ref{eqn.uf})  were first obtained by 
FKS [\onlinecite{kfs}] who considered the interaction between guiding 
centres of coherent states in the $n^{th}$ \la .

The asymptotic expression (\ref{eqn.ufas}) for $\uf$~holds as long as
$p \ll \sqrt{n}$, and
depends on $p$~only via the product $p \: \rla$~indicating that the
important lengthscale for the exchange potential is $\rla$. The
functions $\uh$~and
$-\uf$ for the hardcore potential are plotted in Fig. \ref{f11} for 
$n=20$. The main features are the following. In the range
where $|\uhf |$~is largest, $\uf / \uh \sim n^2$; the absolute minimum
of 
$-\uf$~ is constant (given the rescaling which keeps $vn$ constant)
whereas 
that of $\uh$~vanishes as $n^{-{2\over{3}}}$. Hence, for large $n$,
the Hartree 
potential is insignificant.

Following the analysis by FKS, we expect to find a periodic sequence
of fully occupied and empty stripes of Landau orbital guiding centres,
the 
periodicity of which is given by the prominent first
minimum at $p_{{\rm min}}$~of $-\uf$. With $p_{{\rm min}} \simeq
{2.16/{\rla}}$, the wavelength is approximately $\lambda_{{\rm
min}} \simeq 2.9 \rla$, about
ten percent larger than that found by FKS for the potential
$\vag$. The origin of this difference lies in the fact that our
contact  interaction is less
hostile to density variations on large lengthscales than a Coulomb
interaction. 

Higher harmonics do not destabilise this CDW, because the oscillations
of $\uf$~are incommensurate with $p_{{\rm min}}$, and because the
principal minimum is very pronounced.


\subsection{Ground state in a system of infinite size}  
\label{subsect.stability}
\label{subsect.survey}

A detailed study of Hartree Fock ground states for a system of
electrons interacting via the screened Coulomb potential $\vag$~has
been carried out by FKS. Whilst it is likely that their conclusions
apply to 
a system with the  hardcore potential, we are unable to make
exact statements either about the lowest energy Hartree Fock state or
about the importance of quantum fluctuations around a Hartree Fock
state at zero temperature. 

The Hartree Fock problem is simplified in principle by the result of
section \ref{sect.effhf}: we need to find the Fourier transform of the
electron density, $\nq$, that minimises $\int \Dq \tilde{V}_{{\rm
eff}}({\bf q}) |\nq |^2$, subject to the constraint that $\nq$~arises
from  occupation only
of states within the $n^{th}$~\la. In fact, this constraint is very
hard to incorporate into a rigorous approach. Despite these obstacles,
FKS, studying $\vag$, have proposed very plausible candidates for the
ground state on the basis of persuasive physical arguments, supported
by extensive numerical calculations. They find a unidirectional CDW
ground state for $\nu$~near $1/2$, and a triangular CDW for $\nu$~away
from $1/2$. Our results at finite temperature, described in section
\ref{sect.cdw}, mirror the conclusions
of FKS at zero temperature. 

In addition to the difficulty of showing rigorously whether the
states discovered by FKS lie at the global minimum of the Hartree Fock
energy, there is a second problem in an exact treatment. It is to show
whether quantum fluctuations around the Hartree Fock state are
important. 

As a restricted illustration of this difficulty, we
consider fluctuations starting from a stripe state. Referring to Fig.
\ref{f12}, there are non-vanishing matrix elements
connecting the stripe state to a state slightly higher in energy
with two electrons at the edges moved by an infitesimal distance $(d
\alt 
\lambda_{{\rm F}})$ into the empty regions. For a
system on a finite cylinder, these matrix elements are absent due to
the 
discrete locations for the centres of Landau orbitals.
 For the infinite system, however, this
problem persists, and we address it as follows. The self-energy
$\sik$~(\ref{eqn.selfenergy}) will not be 
changed noticeably by changes in orbital occupation at the edges
of the stripes because their
width is only a fraction  
$n^{-1}$~of the total stripe width, and $\sik$~varies slowly across
the 
edge. 
After linearising $\sik$~around the
edge, the 
excitations of Fig. \ref{f12} acquire a linear dispersion
relation. The problem of one isolated stripe then corresponds  
to a Luttinger liquid [\onlinecite{macdlutt},
\onlinecite{mahan}], where one edge of the stripe provides the
left-moving fermions and the other, the right moving ones. As a
result, sharp edges in orbital occupation are smoothed out
[\onlinecite{mahan}], but only over a fraction of the stripe width
that vanishes for large $n$. It seems likely that this feature of the
behaviour should persist for a system of many stripes and hence that
quantum fluctuations around the stripe state are unimportant for a
contact potential.


\section{Comparison with behaviour for finite range potentials}
\label{sect.finiterange}


\subsection{Diagrammatic analysis}
\label{subsect.nonhf}

If the two-particle interaction has a 
finite range, the diagrammatic analysis described in
appendix \ref{app.dia} does not
simplify in the same way as for the hardcore potential.  To
demonstrate the essential differences, we consider, as an example of a
finite range potential, $V_2$
(\ref{eqn.finiterange}).  For simplicity, we include a 'neutralizing
background' 
so that $\tilde{V}_2(0)=0$.

Consider first the exchange diagram with $k$ interaction lines 
(A in Fig. \ref{f15}). Its contribution to the self energy scales with
$n$ as $\left[ \sum_l
M(0,l) \right]^k \sim \left[ {v_2 R/{\rla}} \right]^k$, where $\rla
\sim 
n^{1\over{2}}$. We show below that 
diagrams with crossed interaction lines (B in Fig. \ref{f15}) or
$k-1$~closed fermion loops (C in Fig. \ref{f15}), 
make a contribution of the same order to
the self-energy. Therefore, these diagrams cannot be discarded  
for large $n$. We emphasize that, having projected the
Hamiltonian onto a single \la, the interaction strength sets the only
energy scale. As previously, we choose this scale to be our unit of
energy, setting ${v_2 R/{\rla}} = 1$.

We now show for the case $k=2$~that all three  diagrams in
Fig. \ref{f15} have the same $n$-dependence for large $n$. We denote
the integrals over intermediate states by $I_A, I_B$~and $I_C$ and
obtain

\bea
I_A & = & \left[ \int q dq w(q) \right] ^2 \nonumber \\ 
I_B & = & \int q dq \left[ w(q) \right] ^2 \nonumber \\
I_C & = & \int \int q_1 dq_1 q_2 dq_2 J_0(q_1 q_2) w(q_1) w(q_2)
\nonumber
\eea
where $w(q)=\tilde{V}(q) \aqt$. It can easily be checked that the main 
contribution to these integrals arises from the region where $q$~is of
order 1. In this region, $w(q) \sim n^{-{1\over{2}}}$. Therefore,
$I_A$~and $I_B$~are of the same order. The Bessel function $J_0(q_1
q_2)$~in $I_C$~oscillates slowly in this region so that $I_C$, also, 
makes a contribution of the same order in $n$.

The analysis can be extended to all $k$. The crucial difference
between between finite range and the hardcore interactions is the
following. While the contribution to 
$I_A, I_B$~and $I_C$~arising from the region $q \sim
1$~does not differ in both cases, the dominant
contribution for the hardcore interaction arises from  $q \gg
1$, and here the dependence on $n$ of $I_A, I_B$~and
$I_C$~differs. By contrast, for finite range potentials,
$\tilde{V}(q)$~is  
falls off for $q \gg R^{-1} \sim 1$, and the large
$q$~region never contributes. Since this feature is common to all
finite range potentials, we expect the above argument to be applicable
generally. In particular, for $\vag$~, assuming that divergences
arising in 
integrals such as in Eq. (\ref{eqn.loopint}) are removed by screening
for  $q \alt n^{-{1\over{2}}}$,
we reach the same conclusions up to logarithmic corrections. 
Hence, the Hartree-Fock approximation for 
finite range potentials is not exact even as \ngi, and 
omits diagrams of the same order in $n$~as those retained.

Nevertheless, as the range of the potential is increased from zero,
corrections to the Hartree-Fock approximation are governed by the
small parameter $R/l$. For the second order diagrams, this can be seen
as follows. Consider first the integrals $I_A$~and $I_C$. The integral
$I_C$~is effectively cut off by the oscillations of the Bessel
function when $q_1 q_2 \gg 1$, whereas the integrand of $I_A$~is not
cut off until $q \sim R^{-1}$, so that $I_C / I_A$~is parametrically
small in $R/l$. Similarly, the contribution of the region $R^{-1} \agt
q \agt 1$~to $I_B$~compared with its contribution to $I_A$~is
parametrically small in $R/l$, and it is from this region that the
dominant contribution to $I_A$~arises.  

The condition $R/l \ll 1$~is fulfilled for the screened Coulomb
interaction, $V_{{\rm AG}}$, for weak
fields: taking the effective range of $V_{{\rm AG}}$~to be $a_B$,
$a_B/l \sim 
1/10$~ in GaAs for a magnetic field B$\sim 70 {\rm mT}$.

One can still ask whether there in fact exist any competitors
to a CDW groundstate even for $R/l \agt 1$. We consider one such
example: the Tao-Thouless 
state at filling fraction $\nu = 1/t$,  
in which one in $t$~Landau orbitals is occupied in a regular
fashion. It is of interest because it is  qualitatively different from
the CDW  states, having
 a uniform charge density, and it is sufficiently simple to allow
detailed calculation. We find the cohesive energy of the Tao-Thouless
state scales as $vn^{-{1\over{2}}}$. For comparison, the cohesive
energies of the CDW groundstates with the interactions $V_2$~and
$\vag$~scale as $v n^{-{1\over{2}}}$.
  The nature of the true groundstate in
high \la s with interactions with range comparable to or larger than
the magnetic length $l$,
appears to be open and may depend on the details of the interaction. 

Tao and Thouless
calculated the cohesive
energy of their state in the random-phase approximation.
The starting point of the self-consistent calculation is as
follows: it is assumed that the energy for an occupied Landau orbital
is 
$\eta_p$~and for an empty orbital is $\eta_h$. This
determines a gap, $\Delta = \eta_h - \eta_p$. The RPA series is then
summed  to obtain 
expressions for $\eta_p$~and $\eta_h$ in terms of $\Delta$. These
equations 
are solved self-consistently. Generalizing the work of
Tao and Thouless [\onlinecite{tt}] to higher \la s and to a
wavevector-dependent dielectric constant yields the following  
equation  

\bea
\Delta & = & \eta_h - \eta_p  \nonumber \\
& = & \int_0^\infty {w(q)^2 q dq
\over{(\Delta ^2 
t ^2 + 2 t \Delta w(q))^{{1\over{2}}} + \Delta t + 2 w(q)}} -
\nonumber\\
&&\int_0^\infty (1- {2\over{t}}) {w(q)^2 q dq \over{\Delta t + 2
w(q)}} .
\label{eqn.del}
\eea

We wish to know the scaling of $\Delta$, and hence of the cohesive
energy, with $n$. 
By considering the behaviour of the integrand as \ngi , assuming
$\Delta = \Delta_0 n^\alpha$ (with $\Delta_0$~independent of $n$),
we find
$\alpha$~ 
simply by comparing powers of $n$ on both sides of
(\ref{eqn.del}). 
We find
that $\alpha = -1/2$ for any fixed $t$.

 Thus,
the cohesive energy of the CDW in the Hartree-Fock
approximation has the same order as the cohesive energy of the uniform
density Tao-Thouless state.


\subsection{Influence of the potential range on the structure of the
charge density waves}
\label{subsect.fate}

Even within the Hartree Fock approximation, there are significant
differences in the structure of the charge density waves formed with
hardcore or finite range interactions.
Near half-filling, FKS find a state of alternating incompressible 
(fully occupied) and
empty stripes of Landau orbitals for small and moderate
$n$. Asymptotically for large $n$, however,  they
find that
 the stripes
break up into narrower stripes. When the guiding centre density of
these narrower stripes is 
coarse-grained, a sinusoidal oscillation in density at wavevector
$p_{{\rm min}}$ results.
  The origin of the break-up of the stripes lies in the fact
that, for \ngi, the energy cost for  higher harmonics in guiding
centre density  outweighs 
the gain due to the fundamental one. This happens because it is
the Hartree potential which sets the energy scale for
higher harmonics, while the scale for the energy of the fundamental
harmonic is set by the Fock potential. At large $n$, the Hartree
potential is parametrically larger than the Fock potential.
 The higher harmonics are  avoided by the break up into narrower
stripes, which establishes a 
quasisinusoidal density distribution. For the hardcore potential,
this does not occur; rather, due to the
vanishing of $\uh$, the energy cost for higher harmonics does not grow
relative to the
fundamental one as \ngi. 

In order to see whether the break-up found by FKS is a general feature
of potentials with 
non-zero range, we investigate the two simple finite range
potentials introduced in Eq. (\ref{eqn.finiterange}).
In both cases, we find that $\tilde{u}_{i{\rm_H}}(q) =
\vft_i(q) \aqt$~remains of order $v_i$~for $q$~of order
$n^{-{1\over{2}}}$ as \ngi. For $\tilde{u}_{i{\rm_F}}$, this is not
the case. The integral $\tilde{u}_{i{\rm_F}}(p) = \int q dq \vft
_i(q) \aqt J_0(pq)$ is effectively cut off  at  $q = R^{-1}$,
and we find for large $n$

\beq
{\uf \over {\uh}} \sim {1 \over{R \rla}} = n^{-{1\over{2}}} 
\label{eqn.finrange}  
\eeq
Since the fall off of $\vft (q)$, and hence of the integral for $\uf$,
above a wavevector $q \sim R^{-1}$ is a feature of all finite range
potentials, we expect the stripes to break up
when $\rla \gg R$. If the effective range of the Coulomb potential
is taken to be the Bohr radius, this condition corresponds to
$r_s^{-1} = \kfe a_{{\rm B}} / \sqrt{2} \gg 1$, which is where FKS
indeed found the destruction of the wide stripes to occur.

We note that, for finite range potentials, there also exist other,
quite different routes by which stripes may become unstable. For
example, for the potential $V_1$, $\uhf$~has minima of comparable
depth at $p \sim n^{-{1\over{2}}}$~and at $p \sim R^{-1}$. Then, it
may be more advantageous to build a correlated state where
correlations on the scale $R$~are as important as or more important
than 
correlations on scales of $\rla$.

Finally, it should be remarked that
the transition to a charge-density wave will become
increasingly hard to observe for \ngi, since the transition
temperature, being proportional to $\tilde{u}_{{\rm F}}$, decreases
with increasing $n$ essentially  
as $n^{-{1\over{2}}}$.


\section{SUMMARY}

In this paper we have considered electrons in a high \la~interacting
via 
a hardcore potential and found that this is a rare example of a
many-body problem in more than one dimension which is,
at least in part, exactly solvable.

The focus of our work is the extent to which it is possible to obtain
mathematically controlled conclusions in the  high \la~limit. This
turns out to be very dependent on the form of the two-particle
interaction potential. For the hardcore potential, we have shown that
Hartree  Fock theory is exact 
and found a transition to both
unidirectional and triangular  charge-density waves at
finite temperatures. For interactions of finite range $R$, this is not
the case. Nonetheless, for short-ranged interactions, $R/l$~is  a
small  parameter governing corrections to the
Hartree-Fock approximation. For $R/l \agt 1$, 
 we argue that there exist uniform density
states which may be competitors with the charge density waves for the
ground state.

In the context of the lowest \la, work with the hardcore potential
has yielded some significant conceptual simplifications while
retaining the essential physical features of the problem
[\onlinecite{macdrev}]. Whether this is also the case in high Landau
levels 
will ultimately have to be settled by experiment.

\begin{acknowledgements}

One of us (JTC) is grateful to L. Glazman and B. I. Shklovskii for
helpful 
discussions and for copies of references [\onlinecite{ag}] and
[\onlinecite{kfs}] prior to publication.
This work was supported in part by EPSRC grant GR/GO 2727.

\end{acknowledgements}


\appendix
\section{Bounds on the contribution from non-Hartree Fock diagrams}
\label{app.dia}

\begin{figure}

\centerline{\psfig{figure=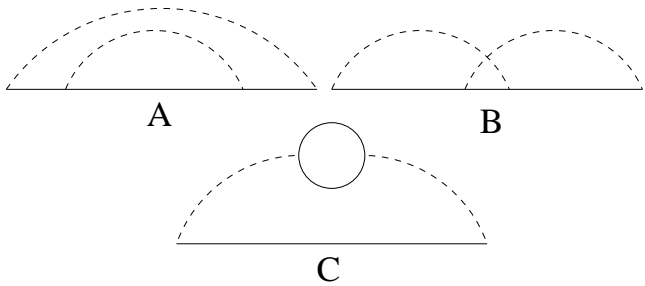,width=8.5cm}}
\vspace{0.5cm}
\caption{The three classes of diagrams.}
\label{f15}
\end{figure}

We compare the $n$-dependence of the three types of diagrams
illustrated in Fig. \ref{f15} for each
order $k$ of the interaction. For the first class of diagrams (A in 
Fig. \ref{f15}), it
follows from Eq. (\ref{eqn.norm}) that their $n$-dependence is 
\beq
(vn)^k
\label{eqn.normk}
\eeq

In the second class of diagrams (B in 
Fig. \ref{f15}), those with crossed interaction lines,
we use an analysis following  work by
Benedict and Chalker [\onlinecite{bc}] on the scattering of electrons
in high \la s with 
disorder. 

\begin{figure}

\centerline{\psfig{figure=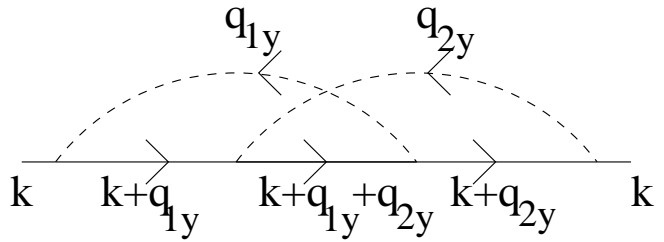,width=8.5cm}}
\vspace{0.5cm}
\caption{Diagram with two crossed interaction lines}
\label{f2}
\end{figure}

Consider the diagram with two crossed lines (Fig. \ref{f2})  which
corresponds to the integral $\int \int
d\qoy d\qty M(\qty,\qoy) M(-\qoy ,\qty)$. From this, we can guess the
general structure for the integral over
intermediate states $\ix$ for
$r$~interaction  lines. It is
\beq
 \ix= \int \prod_{i=1}^r d\qiy M(\Qi,\qiy),
\eeq
where $\Qi = \sum_{j=1}^r Z_{ij} q_{jy}$~with $Z_{ij}=0$~if
the 
interaction lines $i$ and $j$ do not cross and $+1 (-1)$~if they cross
and $i$ starts to the left (right) of $j$. 

We obtain an upper bound on $|\ix |$ by first picking a pair
$t,u$~such that $Z_{tu}=1$~and then relabeling indices such that $t =1
$~and  $u = 2$. Then

\bea
|\ix | & \leq & |\int \prod_{j=3}^r dq_{jy} \iot M(Q_{j},q_{jy})|
\nonumber \\ 
\iot & \equiv & \int \prod_{i=1}^2 d\qiy \: M(a_{1}+\qty,\qoy) \:
M(a_{2}-\qoy,\qty) \nonumber 
\eea

The $a_i$~are independent of $\qoy, \qty$, so that we can use the
Cauchy-Schwarz inequality to obtain:

\bea 
|\iot| & \leq & \int \int  d\qoy d\qty |M(\qoy ,\qty)|^2 
\label{eqn.causchw} 
\eea
\bea
|\ix | &\leq & \int \prod_{i=3}^r d\qiy |M(Q_{i},\qiy)| \int \int
d\qoy  d\qty
|M(\qoy ,\qty)|^2 
\label{eqn.csdecoup} \\
|\ix | & \leq & \int \prod_{i=3}^r d\qiy |M(0,\qiy)| \int \int d\qoy
d\qty 
|M(\qoy ,\qty)|^2
\eea

The last line follows since  $|M(0,l)| \leq |M(m,l)|$~for the
hard-core potential.
The integrals over $\qiy~(i \geq 3)$~ are now decoupled and give the
same  contribution as (\ref{eqn.normk}). The remaining $n$-dependence
is fully contained in $|\iot| \leq (2 \pi)^{-2} \int {q dq} (\vft(q)
\aqt)^2$. For our potential, $(\vft(q)\aqt)$~is monotonically
increasing up to the turning point at $q=\sqrt{8n}$ where it is
proportional to $(vn) n^{-{2\over{3}}}$. Replacing one of the factors
$(\vft(q)\aqt)$~in 
$\iot$~by this upper bound, the remaining integral is given by Eq.
(\ref{eqn.norm}), so that $\iot$,
and hence $\ix$, vanishes at least as $n^{-{2\over{3}}}$~compared to
the exchange diagram with the same number of interaction
lines. Therefore,  all diagrams containing
crossed interaction lines and no fermion loops make a vanishing
contribution 
as $n \rightarrow \infty$. 

\begin{figure}

\centerline{\psfig{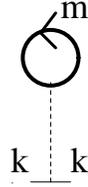}}
\vspace{0.5cm}
\caption{Hartree loop. The solid line is a full propagator.}
\label{f4}
\end{figure}

Next, consider the class of diagrams containing closed fermion
loops (C in Fig. \ref{f15}).  
First examine those which contain Hartree loops (Fig.
\ref{f4}) 
only. A Hartree loop gives rise to an integration of the form 
$\sum_{m} M(m,0)$ = $(2 \pi)^{-2} \int \int {d\qx} dm \vft (\qx)
\alpha_n ({\qx
^2/{2}}) e^{-i\qx m} = 0$, 
since $\vft_{{\rm HC}} (0) = 0$. Similarly,
all other diagrams 
containing Hartree loops vanish for our potential, a fact unrelated 
to the large-$n$~limit.
Note that this is no longer the case in a state with broken
translational symmetry. 

\begin{figure}

\centerline{\psfig{figure=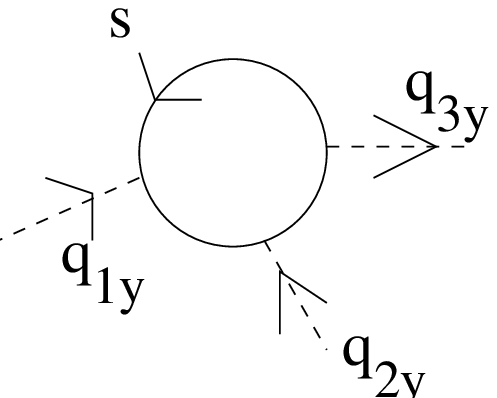,width=4.0cm}}
\vspace{0.5cm}
\caption{Closed fermion loop.}
\label{f3}
\end{figure}

Finally, we turn to those diagrams which contain at least one
non-Hartree fermion loop such as the one in Fig. \ref{f3}. Each of
those loops give rise to an integral over the loop pseudomomentum $s$, 
which leads to a $\delta$-function $\delta
(q_{1x}+q_{2x}-q_{3x})$. Similarly, pseudomomentum conservation gives
rise to a corresponding condition on the $\qiy$. The integral
over  intermediate states, $I_L$ of diagrams containing k
interaction lines and L non-Hartree loops is then bounded above by

\beq
|I_L | \leq |(\prod_{j=1}^{k-L} \int {d{\bf q}_i\over{2
\pi}})(\prod_{j=1}^{k} 
\alpha_n(\Qi ) \vft (\Qi))| 
\label{eqn.loopint}
\eeq 
where ${\bf Q}_i = \sum_{j=1}^{k-l} Y_{ij} {\bf q}_j$ and
$Y_{ij}=0,\pm 1$ is fixed by the constraints imposed by the
$\delta$-functions. This is an upper bound since we have left out the
oscillatory term in the expression for the vertex. 

We can again replace the $|\alpha_n(\Qi ) \vft (\Qi)|$~by an upper
bound and finally obtain $|I_L | \sim ({vn})^k ({{n^{-{2\over{3}}}}})^L
\sim 
n^{-{2L\over{3}}}$. Hence these diagrams also make a vanishing
contribution  as
$n \rightarrow \infty$.


\section{Calculation of the Fock potential}
\label{app.fock}

We start by writing the Fourier transform of the real space density,
$\tilde{n}({\bf q})$, in terms of the density matrix $\rkkp$~in the
Landau orbital basis.

\begin{eqnarray}
\nq & = & \sum_{k,k^{\prime}} \langle k \pri |e \iqr |k
\rangle \rkkp \nonumber \\
    & = & {L\over{2 \pi}} \left( \int dk e^{i k \qx} \rkqy \right)
{\rm exp}({i \qx \qy \over{2}}) \omega_{n}({q^2 \over{2}}) \nonumber \\
    & \equiv & {L\over{2 \pi}}  {\tilde{\rho}}(\qx, \qy)
{\rm exp}({i \qx \qy \over{2}}) \wq
\label{eqn.nrho}
\end{eqnarray}

This can be inverted to give ${\tilde{\rho}}(\qx, \qy)$, and hence
$\rkkp$, in terms of $\nq$.

The Fock potential in terms of the real space density matrix
$\rrrp$~is
\beq
V^{\rm{F}}(\br, \br \pri)  =  V(\br - \br \pri) \rrrp. \nonumber
\eeq

Substituting for $\rkkp$ from (\ref{eqn.nrho}) gives 
\bea
V^{\rm{F}}(\br, \br \pri) & = &
   \sum_{k,\qy} V(\br - \br \pri) \langle \br \pri |k \rangle
\langle k+\qy | \br \rangle \rkqy
\eea
and
\bea \langle l|V^{\rm{F}}|m \rangle & = &  \sum_{k, \qy}
\int {\Dp\over{(2 \pi)^2}} \langle l|e^{i{\bf{p \cdot r}}}|k \rangle
\times \nonumber \\
&& \langle k+\qy|e^{-i{\bf{p \cdot r}}}|m \rangle \rkqy
{\tilde{V}}({\bf{p}}) 
\label{eqn.lvm1} \\   
  & = &  \int {\Dq \over{(2 \pi)^2}} \nq { \langle l|e^{-i{\bf{q \cdot
r}}}|m \rangle} {\tilde{V}}_{{\rm{eff}}}^{{\rm F}}({\bf{q}}),
\label{eqn.lvm2}  
\eea
where the Fock potential is
\beq
  {\tilde{V}}^{{\rm{F}}}_{{\rm{eff}}}({\bf q})  = 
 {1 \over{\aq}} \int
{\Dp \over{2 \pi}} {\tilde{V}}({\bf{p}}) e^{-i{\bf{p \cdot
q}}} \ape.
\eeq

\begin{figure}

\centerline{\psfig{figure=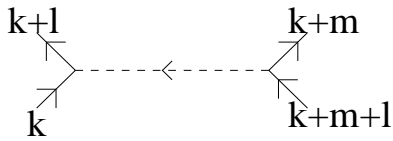,width=5.0cm}}
\vspace{0.5cm}
\caption{Interaction vertex $\mml$.}
\label{f1}
\end{figure}

\vspace{0.5cm}



\vspace{0.5cm}

\begin{figure}

\centerline{\psfig{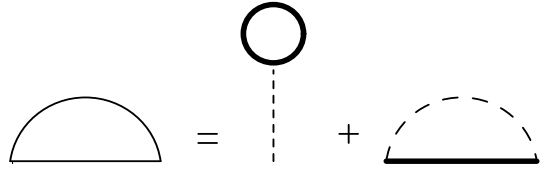}}
\vspace{0.5cm}
\caption{Dyson equation for the self-energy.}
\label{f5}
\end{figure}
\vspace{0.5cm}

\begin{figure}

\centerline{\psfig{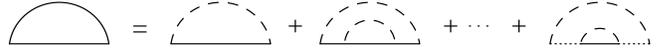}}
\vspace{0.5cm}
\caption{The self energy for the high-temperature phase.}
\label{f6}
\end{figure}
\vspace{0.5cm}

\begin{figure}

\centerline{\psfig{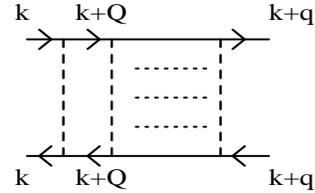}}
\vspace{0.5cm}
\caption{Ladder diagrams for the two-particle propagator. The solid
lines 
denote full propagators.}
\label{f7}
\end{figure}
\vspace{0.5cm}

\begin{figure}

\centerline{\psfig{figure=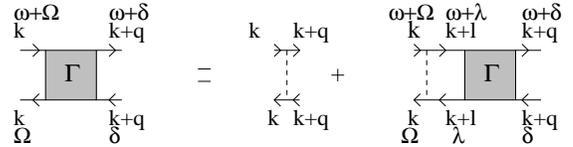,width=7.0cm}}
\vspace{0.5cm}
\caption{Bethe-Salpeter equation for $\Gamma$. Greek letters denote 
Matsubara frequencies.}
\label{f8}
\end{figure}
\vspace{-2.0cm}

\begin{figure}

\centerline{\psfig{figure=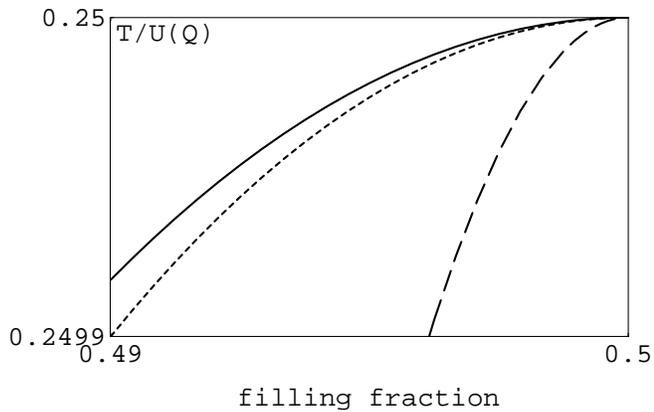,width=8.5cm}}
\vspace{-3.0cm}
\caption{Phases near the triple point. The solid line,
$T_1$, separates the uniform and triangular phases, the short dashes, 
$T_2$, are the spinodal line and the long dashes, $T_{12}$, separate
the  
triangular and unidirectional phase.}
\label{f13}
\end{figure}
\vspace{-2.0cm}

\begin{figure}

\centerline{\psfig{figure=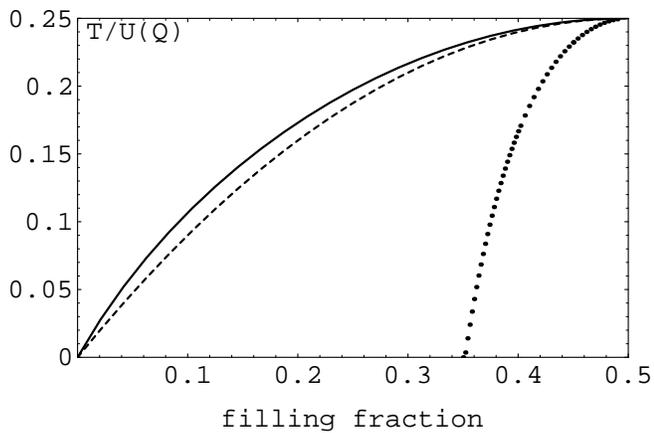,width=8.5cm}}
\vspace{-3.0cm}
\caption{Phase diagram as in the previous figure. The dotted 
line is an extrapolation of $T_{12}$.}
\label{f14}
\end{figure}
\vspace{-2.0cm}

\begin{figure}

\centerline{\psfig{figure=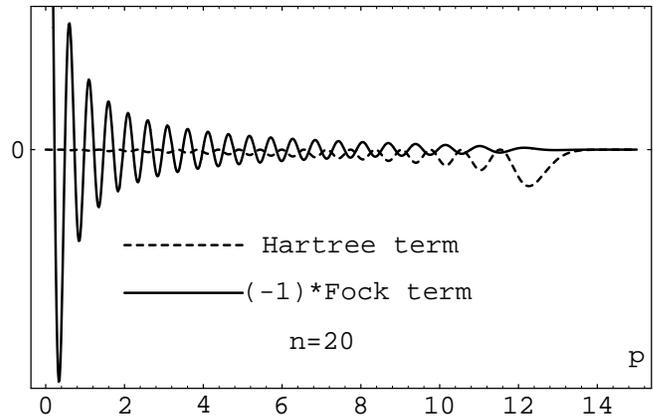,width=8.5cm}}
\vspace{-2.0cm}
\caption{$\uh$~and $-\uf$. The unit of $p$ is the inverse magnetic
length; the unit of energy is arbitrary.}
\label{f11}
\end{figure}
\vspace{0.5cm}

\begin{figure}

\centerline{\psfig{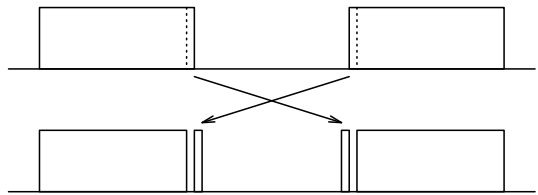}}
\vspace{0.5cm}
\caption{Two near degenerate states $| A \rangle, |B \rangle$~with 
nonvanishing matrix element $\langle A |V_{{\rm HC}}| B \rangle$.}
\label{f12}
\end{figure}
\vspace{0.5cm}



\vspace{0.5cm}

\vspace{0.5cm}


\begin{references}

\bibitem{fqhe}
{\em The Quantum Hall effect}, edited by R. E. Prange and S. M. Girvin
(Springer-Verlag, New York, 1990), 2nd ed.

\bibitem{brout} 
R. Brout, Phys. Rev. {\bf 118}, 1009 (1960)

\bibitem{voll} 
W. Metzner and D. Vollhardt, Phys. Rev. Lett. {\bf 62}, 324 (1989)

\bibitem{old} 
see, for example, A. H. MacDonald, Phys. Rev. B. {\bf 30}, 3550 (1984)
and N. d'Ambrumenil 
and A. M. Reynolds, J. Phys. C {\bf 21}, 119 (1988)

\bibitem{ag} 
I. L. Aleiner and L. I. Glazman, Phys. Rev. B. {\bf 52}, 11296 (1995)

\bibitem{kfs}
M. M. Fogler, A. A. Koulakov, and B. I. Shklovskii, preprint
cond-mat/9601110 and A. A. Koulakov, M. M. Fogler and
B. I. Shklovskii, Phys. Rev. Lett. {\bf 76}, 499 (1996)

\bibitem{fpa} 
H. Fukuyama, P. M. Platzmann and P. W. Anderson, Phys. Rev. B. {\bf
19}, 5211 (1979)

\bibitem{rs}
M. E. Raikh and T. V. Shahbazyan, Phys. Rev. B. {\bf 47}, 1522 (1993) 

\bibitem{macdrev}
A. H. MacDonald, in {\em Les Houches, Session LXI, 1994, Physique
Quantinque Mesoscopique}, edited by E. Akkermans, G. Montambeaux, and
J. L. Pichard (Elsevier, Amsterdam, 1995)

\bibitem{lau}
R. B. Laughlin, Phys. Rev. Lett. {\bf 51}, 605 (1983) 

\bibitem{tk}
S. A. Trugman and S. Kivelson, Phys. Rev. B. {\bf 31}, 5280 (1985) 

\bibitem{bo}
C. M. Bender and S. A. Orszag, {\em Advanced Mathematical Methods for
Scientists and Engineers}, (McGraw-Hill, New York,
1978), ch. 10.

\bibitem{erd}
A. Erdelyi, J. Indian Math. Soc. {\bf 24}, 235. 

\bibitem{turning}
 It is straightforward to check
that these conclusions are unaltered if one uses the more accurate
asymptotic forms (\ref{eqn.wkbturnz}) and (\ref{eqn.wkbturnn}) near
the WKB turning points.


\bibitem{rh} 
E. H. Rezayi and F. D. M. Haldane, Phys. Rev. B. {\bf 50}, 17199 
(1994) 

\bibitem{tt} 
R. Tao and D. J. Thouless, Phys. Rev. B. {\bf 28}, 1142 (1983)

\bibitem{ladder} 
Actually, another set of ladder diagrams has to be summed as well,
namely 
those in Fig. \ref{f7} rotated by 90 degrees, so that the propagators
run vertically and the interaction lines horizontally. In the
expression for $\tilde{\Gamma}$~these graphs make a vanishing
contribution in the thermodynamic limit. 

\bibitem{yf}
 D. Yoshioka and H. Fukuyama, J. Phys. Soc. Japan {\bf
47}, 394 (1979), D. Yoshioka and P.A. Lee, Phys. Rev. B {\bf 27},
4986(1983) and references therein. R. R. Gerhardts and Y. Kuramoto,
Z. Phys. B {\bf 44}, 301 (1981)


\bibitem{cont}
We do not consider here correlated, large-scale fluctuations that
might result, for example, in defect-mediated melting of the CDW.

\bibitem{mr} 
A. Malaspinas and T. M. Rice, Phys. Konden. Mater. {\bf
13}, 193 (1971) and K. Nakanishi and K. Maki, Progr. Theor Phys. {\bf
48}, 1059 (1972)

\bibitem{mg} 
A. H. MacDonald and S. M. Girvin, Phys. Rev. B {\bf 38}, 6295 (1988)

\bibitem{macdlutt}
A. H. MacDonald, preprint cond-mat 9512043

\bibitem{mahan}
G. D. Mahan,  {\em Many-Particle Physics}, (Plenum Press, New York,
1990), ch. 4.4.

\bibitem{bc}
K. A. Benedict and J. T. Chalker, J. Phys. C {\bf 19}, 3587 (1986).

\end{references}
\end{document}